\documentclass{article}

\usepackage{mathpple}
\usepackage{graphicx}
\usepackage{url}
\usepackage{paralist}

\usepackage{geometry}
\begin{document}

\noindent
\textbf{Preprint of:}\\
Simon Parkin, Gregor Kn\"{o}ner, Timo A. Nieminen,
Norman R. Heckenberg and Halina Rubinsztein-Dunlop\\
``Measurement of the total optical angular momentum
transfer in optical tweezers''\\
\textit{Optics Express} \textbf{14}(15) 6963--6970 (2006).\\
Online at: \url{http://www.opticsinfobase.org/abstract.cfm?URI=oe-14-15-6963}

\hrulefill

\begin{center}

\Large
\textbf{Measurement of the total optical angular momentum transfer\\
 in optical tweezers}

\vspace{2mm}

\large
Simon Parkin, Gregor Kn\"{o}ner, Timo A. Nieminen, Norman R. Heckenberg\\
and Halina Rubinsztein-Dunlop

\vspace{1mm}

\normalsize
\textit{Centre for Biophotonics and Laser Science, School of
Physical Sciences,\\ The University of Queensland, QLD 4072,
Australia}

\end{center}

\begin{abstract}
We describe a way to determine the total angular momentum, both
spin and orbital, transferred to a particle trapped in optical
tweezers. As an example an LG$_{02}$ mode of a laser beam with
varying degrees of circular polarisation is used to trap and
rotate an elongated particle with a well defined geometry. The
method successfully estimates the total optical torque applied to
the particle. For this technique, there is no need to measure the
viscous drag on the particle, as it is an optical measurement.
Therefore, knowledge of the particle's size and shape, as well as
the fluid's viscosity, is not required.
\end{abstract}

\section{Introduction}
The major application of optical tweezers, since their invention
in 1986 \cite{ashkin86}, has been the study of microscopic
biological systems. The first measurements were made on bacterial
flagella and the transport of organelles within a cell
\cite{block1989,ashkin1990nat}. Advances in force measurement
technology enabled high temporal and spatial resolution force
measurements, for example, on motor proteins walking along
microtubules \cite{svoboda1993nat,finer1994}, polymerase
transcribing DNA \cite{yin1995sci}, protein folding
\cite{kellermayer1997,tskhovrebova1997} and viruses packaging DNA
\cite{smith2001nat}. Optical tweezers force measurements have
provided a great insight into the physics of microscopic
biological mechanisms and this application will continue due to
the huge number of biological systems available for study. To
further our understanding of the microscopic biological world,
different quantitative measurement techniques need to be
developed. Here we present a new technique to measure torques
applied by optical tweezers opening a way to new quantitative
studies of microscopic biological systems and their rotational
dynamics.

Techniques to apply torques using optical tweezers by the transfer
of angular momentum from the beam to the particle are well
established. Absorption is the simplest transfer mechanism,
whereby the particle absorbs the light's angular momentum. Both
spin and orbital angular momentum have been transferred in this
way \cite{he95,friese1996pra}. The first is due to the light's
polarisation, while the orbital component is associated with the
spatial distribution of the light's wavefront. A more elegant
approach, that avoids unwanted heating due to absorption, is to
trap a birefringent crystal with a circularly polarised beam
\cite{friese98nat}. This allows for efficient angular momentum
transfer and the applied torque can be measured optically
\cite{nieminen2001}. The advantage of an optical measurement of
the torque is that knowledge of the properties of particle and its
surrounding environment is not required. The particle's exact
shape or the fluid's viscosity and refractive index do not affect
the measurement of the applied torque. Based on this technique a
micro-viscometer has been demonstrated using a spherical
birefringent crystal as the object which was trapped and rotated
\cite{bishop04}. However such crystals are somewhat difficult to
produce and they are unstable in harsh media, such as acidic
solutions. Therefore it would be useful to quantify torque
transfer via orbital angular momentum as a larger torque
efficiency is available \cite{allen92} and suitable optically
asymmetric objects are more abundant \cite{bonin02,bishop03}. Such
objects include photopolymerised structures with sub-micron
features which have been shown to function as light driven
micromachines \cite{galadja2001}. The ability to
\textit{optically} measure the torque applied to these
micromachines would increase their functionality and would allow
for quantitative measurements of rotational dynamics and feedback
control.

Measurement of orbital angular momentum has been of interest in
the field of quantum information and computing. Computer generated
holograms have been used to measure the orbital angular momentum
of single photons produced by parametric down-conversion. The
experiments showed that orbital angular momentum is conserved
during down conversion and that the orbital angular momentum
states are entangled \cite{mair01}. Measurement of the orbital
angular momentum of an arbitrary beam using a similar technique
has been demonstrated \cite{parkin04}. In these experiments the
torque applied to a elongated phase object by a paraxial laser
beam was determined by measuring the power in the forward
scattered azimuthal modes. However to measure a large number of
azimuthal modes requires a complicated setup and is impractical.
This is the case for measurements involving optical tweezers in
particular, due to the variety of modes that are forward scattered
by a particle trapped in a highly converging and diverging beam.
Radial modes present a further complication for the measurement of
the azimuthal modes.

Optical angular momentum can also be quantified by measuring the
rotational Doppler shift introduced by rotating optical elements
in the beam path \cite{courtial1998}. The orbital component has
been determined by measuring the frequency shift between two beams
with different azimuthal indices \cite{basistiy2003}. In this
experiment one of the beams is rotated using a $90^\circ$ prism.
In principle this technique could be used to measure the orbital
angular momentum of an arbitrary beam. However it would be
difficult to distinguish the different frequencies and, in
particular, their amplitudes due to the noise from the rotating
optical element required to give the frequency shift. A similar
technique based on a Mach--Zehnder interferometer allows sorting
of photons with different angular momentum states
\cite{leach2004}. However, in practice this technique would also
be difficult to implement in optical tweezers due to the series of
interferometers required to measure the angular momentum of an
arbitrary beam.

In this paper we present a method which enables the measurement of
the total optical angular momentum transfer in optical tweezers.
The spin component of the angular momentum can be measured purely
optically as previously \cite{bishop03}. In order to measure the
orbital component we take advantage of this existing method, and
the linear relationship between the rotation rate of the trapped
particle and the torque applied to it, to measure the orbital
angular momentum transferred to the particle. To demonstrate this
method we measure the torque applied to a trapped elongated
particle with a well defined size and geometry. The accuracy of
the particle's dimensions is important in order that the measured
torque can be compared to the viscous drag on the particle as a
test of the validity of the experimental method, but is not
required for the torque measurement itself.

\section{Theory}
A convenient method to optically measure the torque applied to an
object by the spin angular momentum of a laser beam was described
in detail in \cite{nieminen2001}. The torque on the particle due
to spin angular momentum transfer is given by
$\tau_{\mathrm{spin}} = \Delta \sigma P/\omega$
\label{eqn:torquespin} where $\Delta \sigma$ is the change in the
degree of circular polarisation as the beam passes through the
particle, $P$ is the laser power and $\omega$ is the optical
angular frequency.

For a particle rotating steadily the optically applied torque is
equal to the drag torque applied by the surrounding liquid. In a
low-Reynolds-number Newtonian fluid the drag torque is
proportional to the rotation rate. The applied torque is also
equal to the sum of the torques applied by the spin and orbital
components. Therefore the torque acting on the trapped particle is
given by:
\begin{equation}
\tau_{\mathrm{total}} = \tau_{\mathrm{orbital}} +
\tau_{\mathrm{spin}} = \Omega K \label{eqn:torquerotation}
\end{equation}
where $K$ is an unknown constant of proportionality and $\Omega$
is the rotation rate. We assume that the the laser's frequency and
power are known, which means $\tau_{\mathrm{spin}}$ can be found
directly from the change in polarisation ($\Delta\sigma$). This
leaves an equation with two unknowns. By varying the independent
variable ($\tau_{\mathrm{spin}}$) and measuring the rotation rate
($\Omega$), $K$ and the torque due to orbital angular momentum
transfer ($\tau_{\mathrm{orbital}}$) can be determined.

\section{Experiment}

\begin{figure}[!tb]
\begin{center}
\includegraphics[width = 0.8\textwidth]{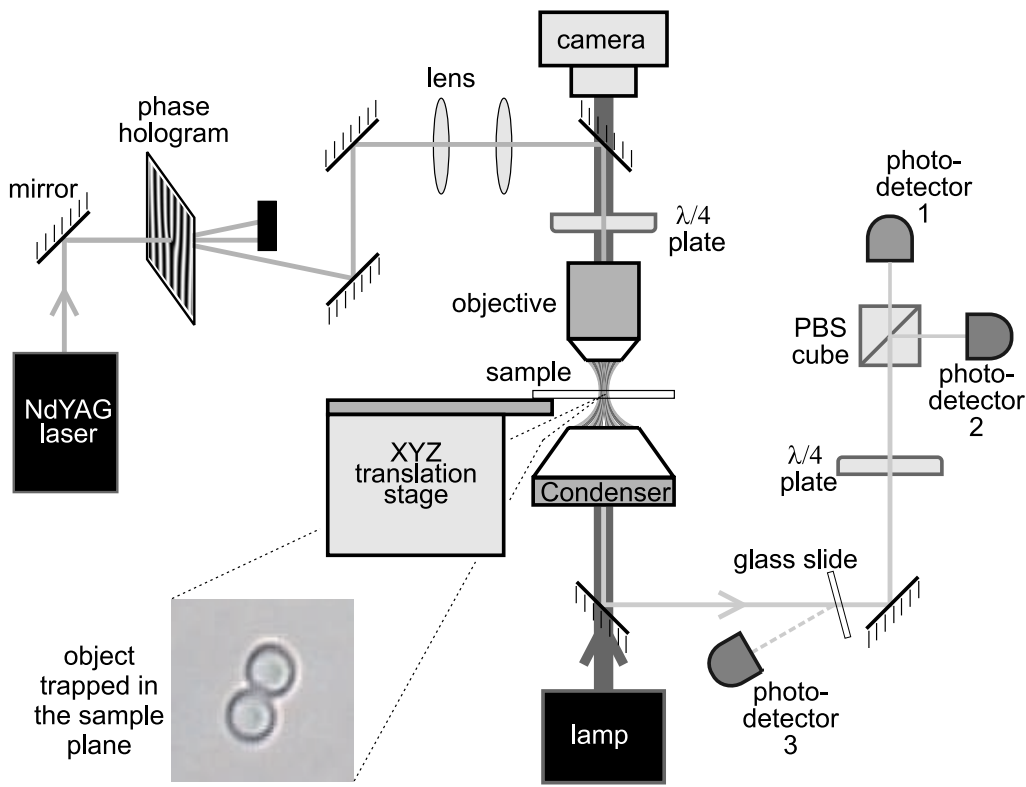}
\end{center}
\caption{The optical tweezers setup used to make measurements of
optical torque applied to a trapped object. The two polystyrene
spheres that were trapped and rotated are shown in the inset. The
phase hologram created a LG$_{02}$ beam in the first diffraction
order, which was used as the trapping beam. The three detectors
measured the rotation rate of the particle and the change in
polarisation of the trapping beam.} \label{fig:tweezersetup}
\end{figure}

The optical tweezers setup (Fig. \ref{fig:tweezersetup}) was used
to make measurements of the optical angular momentum transferred
to trapped particles. A computer generated hologram
\cite{parkin04} was used to generate a Laguerre--Gauss mode with
an azimuthal index of two (LG$_{02}$) in the first diffraction
order. The beam was expanded to fill the back aperture of an
100$\times$ oil immersion objective (NA = 1.3) to yield the best
trapping geometry and efficiency. The quarter wave plate before
the objective was rotated to different angles to make left handed,
right handed or linearly polarised laser light. An oil immersion
condenser, with a numerical aperture greater than the objective
(NA = 1.4), collected the diverging light from the optical trap
created by the objective. A glass slide deflected a small
percentage of the collected light to a photo-detector
(photo-detector 3). The angle between the face of the glass slide
and the axis of the laser beam's direction of propagation was as
close to $90^\circ$ as possible (the angle in the figure is
exaggerated for clarity). This minimised both the amount of light
deflected and the tendency for a certain polarisation to be
deflected more strongly than its orthogonal counterpart. The spot
size of the laser beam was greater than the area of the
photo-detector so that the intensity of only a section of the beam
was measured. The laser light transmitted through the glass slide
was sent to a circular polarisation detection system. The spot
size of the laser light incident on the two detectors (1 \& 2) was
smaller than the detector area so that the two detectors collect
all the laser light collected by the condenser. The polarising
beam splitter cube ensures that the two photo-detectors measure
orthogonal linearly polarised components of the laser light. The
quarter wave plate in front of the polarising beam splitter cube
is aligned to ensure that right circularly polarised light
incident on the quarter wave plate is sent to one detector while
left circularly polarised light is sent to the other detector.

A demonstration experiment was carried out in the optical tweezers
with a simple asymmetric object that is readily available and has
a relatively simple geometry. Two polystyrene beads (each two
microns in diameter) were trapped and pushed together in the
LG$_{02}$ beam so that they behaved as one elongated object.
Although in principle this object could be three dimensionally
trapped, we chose to trap the beads two dimensionally against the
glass slide to overcome the tendency for the elongated dimension
of the object to align vertically in the trap.

Measurement of the laser power at the focus of the objective was
needed for the determination of the optically applied torque on
the trapped particle . A direct measurement is difficult due to
the short working distance and high numerical aperture of the
objective. Therefore the power at the focus was estimated by
determining the transmission of the objective and condenser. As
shown earlier, the optically applied torque also depends on the
change in polarisation of the trapping beam. The circular
polarisation detection system measures the power in each of the
orthogonal circularly polarised components which allows the degree
of circular polarisation to be found.

In order to determine the torque applied by the orbital component
of the beam, the rotation rate of the trapped particles was
measured. The fluctuations in intensity measured by photo-detector
3 corresponds to the rotation rate of the trapped particles. The
particles have two fold symmetry which means that half the
frequency of the intensity fluctuations is the particles' rotation
rate. The constant of proportionality ($K$) between the rotation
rate ($\Omega$) and the optically applied torque
($\tau_{\mathrm{total}}$) can not be measured and was instead
found by making three measurements of $\Omega$ and
$\tau_{\mathrm{spin}}$ at three different polarisations. Although
two polarisations would be sufficient, linear, right handed
circular and left handed circular polarisations were used.

\section{Results and discussion}
\begin{figure}
\begin{center}
\begin{tabular}{ll}
(a) & (b) \\
\hspace{-0.3cm} \raisebox{-0.05cm} {\includegraphics[width =
0.48\textwidth,angle=0]{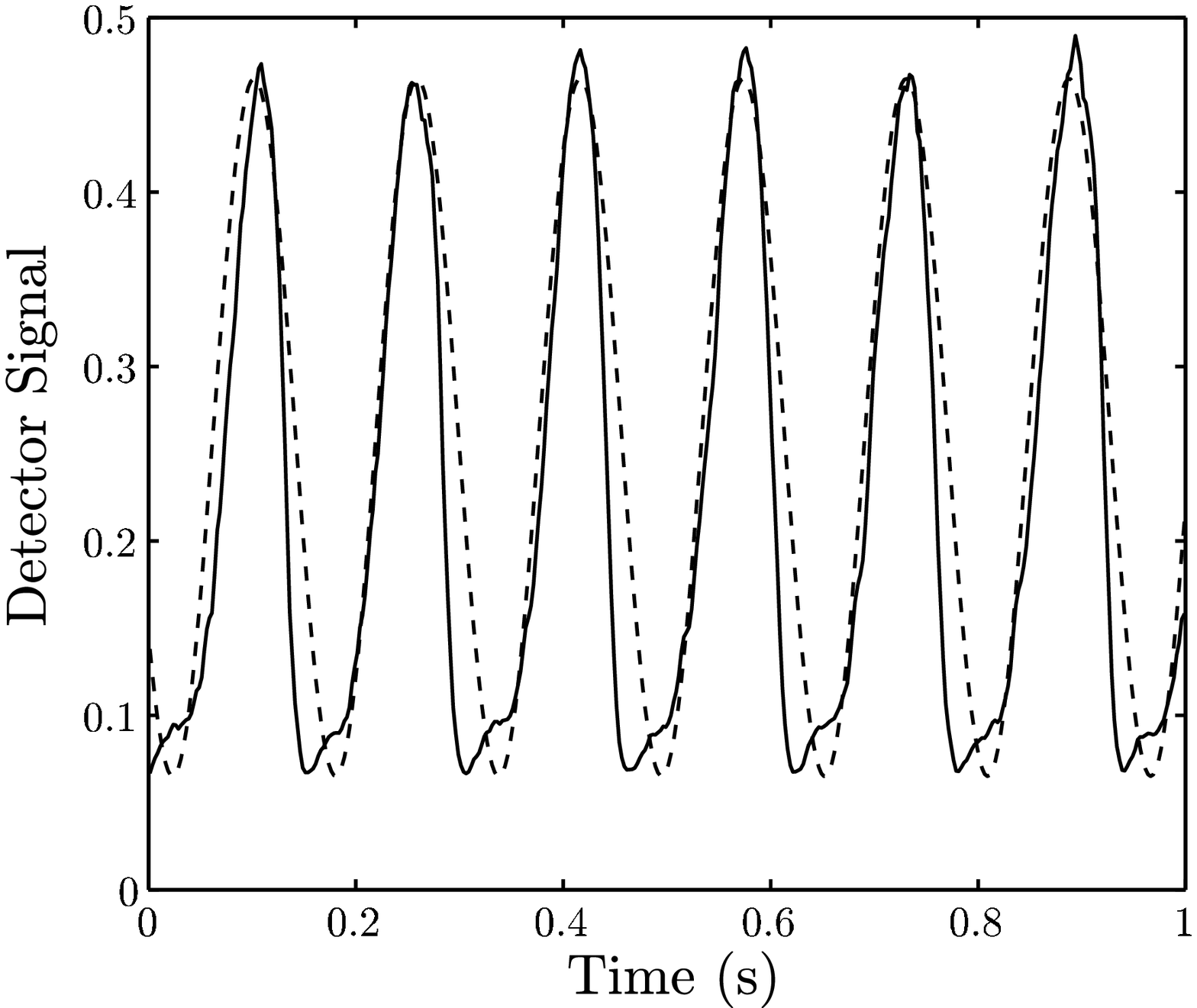}} & \hspace{-0.5cm}
\includegraphics[width = 0.48\textwidth,angle=0]{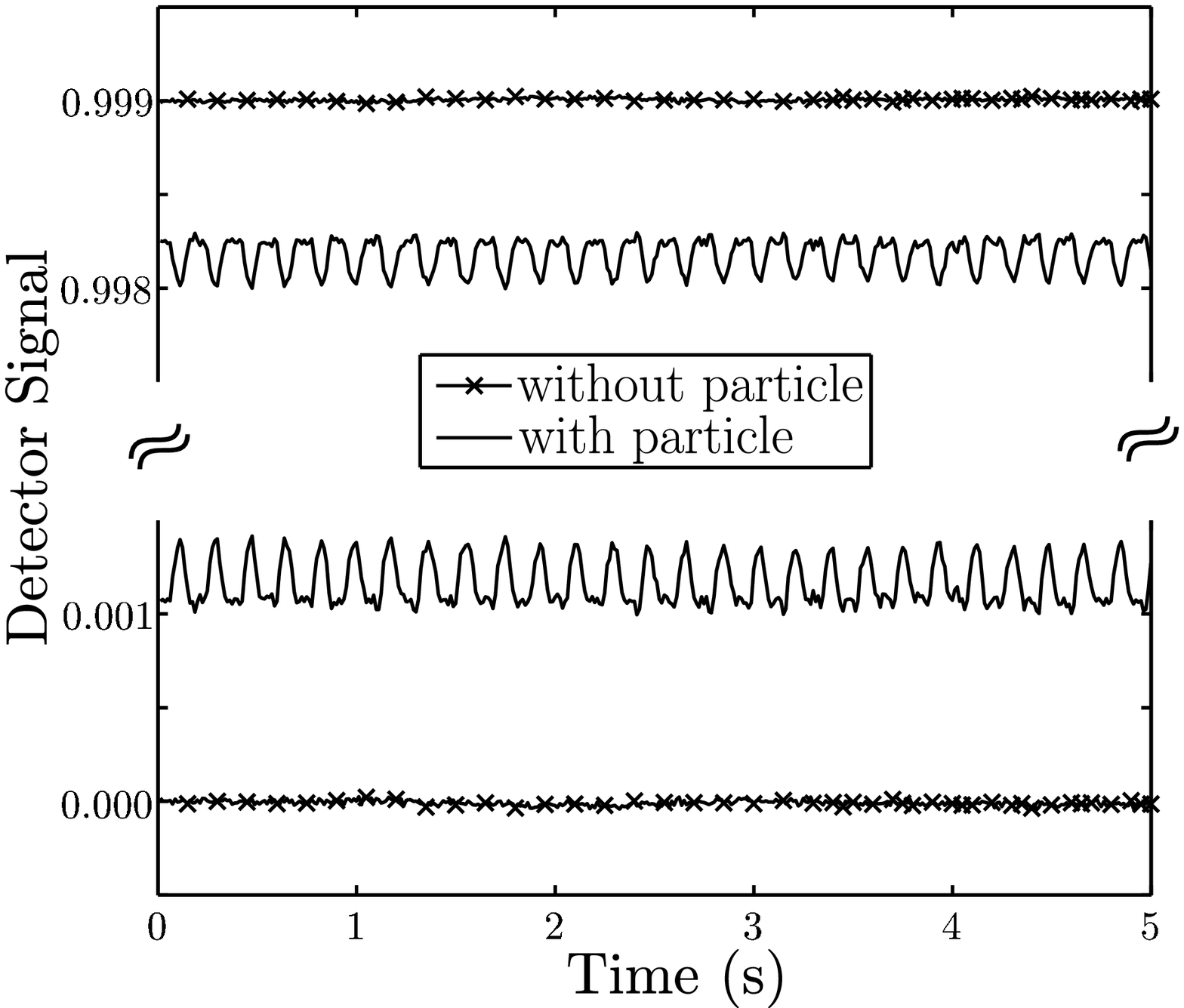}\\
\end{tabular}
\end{center}
\caption{(a) Typical signal from photo-detector 3, from which the
rotation rate of the particle was found. The solid line (---) is
the actual signal from the detector and the dashed line ($--$) is
a sinusoidal fit of the data. (b) Typical signals from
photo-detectors 1 and 2 for the case when no particle is trapped
and also the case when the particles are trapped by the optical
tweezers. The `ripple' visible in the signal with the particle
trapped is due to the rotation of the particle in the trap and has
the same frequency as measured by photo-detector 3. }
\label{fig:rotationrate}
\end{figure}

The change in polarisation of laser light that causes the rotation
of two 2\,$\mu$m beads, as well as their rotation rate, was
measured for each of the three polarisations of the incident beam.
A typical signal from the photo-detector that measures the
particles' rotation rate is shown in Fig.
\ref{fig:rotationrate}(a). A sinusoidal fit to the signal shows
that a constant rotation rate is maintained for a one second
interval. The sampling duration was 5 seconds for each measurement
and the average rotation rate over this interval was found.

The circularly polarised components of the beam were measured by 2
photodetectors. Typical signals from these detectors are shown in
Fig. \ref{fig:rotationrate}(b). The signals for each detector with
and without a particle are plotted for the 5 second sampling
interval. The average for each signal, over the time interval, was
found. The degree of circular polarisation is calculated from
these averages and the change in polarisation ($\Delta\sigma$) was
found by subtracting the signal without a particle from the signal
with a trapped particle.

\begin{figure}
\begin{center}
\includegraphics[width = 0.6\textwidth, angle=0]{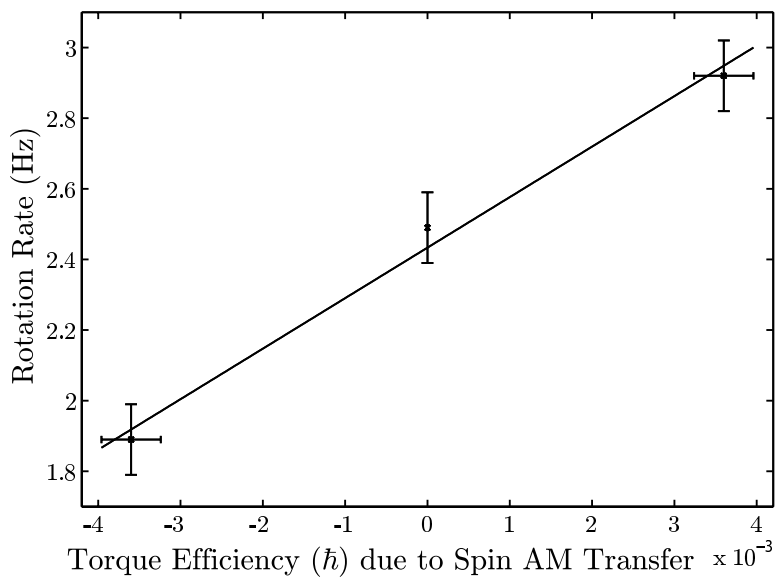}
\end{center}
\caption{The rotation rate, $\Omega$, of the trapped particle (two
polystrene spheres) as a function of the optically applied torque
due to the spin angular momentum of the trapping beam,
$\tau_{\mathrm{spin}}$. The torque transfer of the orbital
component is found from the slope ($1/K$) and intercept
($\Omega_0$) of the fit.} \label{fig:rotationVtorque}
\end{figure}

The rotation rates measured for the three different polarisations
are shown in Fig. \ref{fig:rotationVtorque}. Each point on the
plot has been averaged over 10 data sets. The torque transferred
due to the spin component of the optical angular momentum (on the
\textit{x} axis of the plot) was calculated from the change in
polarisation. The torque transfer via orbital angular momentum is
found directly from this plot using equation
\ref{eqn:torquerotation} and setting $\tau_{\mathrm{spin}}$ to
zero:
\begin{equation}
\tau_{\mathrm{orbital}}=K\Omega_0=\frac{\Omega_0}{\mathrm{slope}}
= 0.017 \pm 0.003 \hbar.
\end{equation}
which means the torque due to the orbital component in this case
is 5 times the torque due to the spin component. It should be
noted here that this is an optical measurement of the torque and
was made without knowledge of the particle's shape or refractive
index, or the surrounding fluid's viscosity or refractive index.
However, in this case, we do know the shape and size of the
particle and the viscosity of the fluid, so the optical
measurement can be compared to a viscous drag model.

For a steadily rotating particle the optically applied torque is
equal to the viscous drag torque on the trapped particles.
Therefore in order to check the measured value for the optically
applied torque, we can model the viscous drag on the rotating
particle. A simple model based on Stokes' drag on a translating
sphere gives the following torque for two adjacent spheres
rotating about their point of contact:
\begin{equation}
\tau_D = 12 \pi \eta a^3 \Omega
\end{equation}
here $a$ is the radius of each sphere (and the lever arm), $\eta$
is the viscosity of the surrounding fluid and $\Omega$ is the
rotation rate of the two spheres. The parameters used for this
calculation are power (20\,mW), rotation rate (2.4\,Hz) and the
individual spherical beads' radius (1\,$\mu$m). The model gives a
torque efficiency of 0.05\,$\hbar$. We estimate the error in the
model could be as much as 50\,\% due to wall effects and
slipstreaming.

The results of the experiment and theory are of the same order of
magnitude. This level of agreement is significant because it
demonstrates that orbital torque can be estimated using this
technique. The difficulty in the presented method is that a small
change in polarisation signal needs to be measured in order to
determine the orbital torque. For an orbital torque as small as
$0.017 \hbar$, the signal measured by photo-detectors 1 and 2 only
changed by $0.2\%$. This level of precision requires careful
alignment of polarising optics and accurate reading from the
photodetectors. The ideal solution to this problem would be to
boost the change in polarisation signal by choosing a particle
that exhibits a stronger birefringence or form birefringence. Such
a particle would allow for an accurate torque \emph{measurement}.
The method, as it stands, has only provided a good \emph{estimate}
of the optical torque transfer in optical tweezers. Fortunately
techniques exist, such as photopolymerisation \cite{galadja2001},
which allow particles with sub-micron features to be fabricated.
Such fine features can enhance the particles form birefringence
which would make them very suitable for torque measurements.

The method described in this paper has the potential to make
accurate measurements of the total optical angular momentum
transfer in optical tweezers when there is negligible absorption.
For proper application of the method a linear dependence of
rotation rate on applied torque is assumed. This means the
particle must freely rotate and that the behaviour of the fluid
must be Newtonian. However, in the case of a particle that is not
freely rotating, for example a particle attached to a cell, the
calibration of the torque can be carried out prior to attaching
the particle. Non-Newtonian behaviours, such as shear thinning are
unlikely to affect the validity of the technique. Firstly because
the shear rates created by particles in optical traps are usually
too small for non-linearities to be observed \cite{knoener2005}
and secondly the trapping power can be reduced to decrease the
rotation rate so that a linear regime of the fluid can be
accessed. Once the torque transfer is calibrated then
non-linearities could be studied.

Feasible applications for this technique are in studying
microscopic biological systems and in the implementation of
micromachines. For example previous studies on bacteria flagella
\cite{block1989,ryu2000} could be extended by measuring the torque
directly. Other applications include torsional elasticity
measurements of single polymer chains or DNA strands as well as
microviscosity measurements \cite{galadja2001}.

\section{Conclusion}
We have demonstrated a technique to measure both the spin and
orbital components of the angular momentum transferred to a
particle trapped in optical tweezers. For this technique, there is
no need to measure the viscous drag on the particle, as it is an
optical measurement. Therefore, knowledge of the particle's size
and shape, as well as the fluid's viscosity, is not required. The
technique successfully estimated the total optical torque on an
asymmetric object when compared with a simple theoretical model.
We have suggested that the accuracy of this method could be
improved by using photo-polymerisation techniques to `tailor-make'
form birefringent particles. Such particles would have the
advantage that both spin and orbital angular momentum could be
more efficiently transferred from the trapping beam. In our
experiment the orbital angular momentum was 5 times the magnitude
of the spin component in this system which suggests that orbital
angular momentum transfer will prove to be useful for quantitative
study of torques in microscopic biological systems and in
micromachine applications.

\section*{Acknowledgements}
We would like to acknowledge the support of The University of Queensland and the Australian
Research Council.

%\bibliography{../../Bibliography/SimonPhD}
%\bibliographystyle{osa}

\end{document}